\documentclass{ws-ijgmmp}

\usepackage{graphicx}  
\usepackage{latexsym}
\usepackage{amssymb}
\usepackage{amsmath}

\newcommand{\beq}{\begin{equation}}
\newcommand{\eeq}{\end{equation}}
\newcommand{\bea}{\begin{eqnarray}}
\newcommand{\eea}{\end{eqnarray}}
\def\eqref#1{(\ref{#1})}

\begin{document}

\markboth{D. Bini, A. Geralico, R.T. Jantzen}
{Petrov type I spacetime curvature: principal null vector spanning dimension}

%
\catchline{}{}{}{}{}
%

\title{\uppercase{ 
Petrov type I spacetime curvature: principal null vector spanning dimension}
}

\author{\footnotesize DONATO BINI}

\address{
Istituto per le Applicazioni del Calcolo ``M. Picone,'' CNR, I--00185 Rome, Italy\\
INFN - Sezione di Roma III, Rome, Italy\\
\email{donato.bini@gmail.com}
}

\author{\footnotesize ANDREA GERALICO}

\address{
Istituto per le Applicazioni del Calcolo ``M. Picone,'' CNR, I--00185 Rome, Italy\\
\email{andrea.geralico@gmail.com}
}

\author{\footnotesize ROBERT T. JANTZEN}

\address{
Department of Mathematics and Statistics, Villanova University, Villanova, PA 19085, USA\\
\email{robert.jantzen@villanova.edu}
}

\maketitle

\begin{history}
\received{(Day Month Year)}
\revised{(Day Month Year)}
\end{history}

\begin{abstract}
The class of Petrov type I curvature tensors is further divided into those for which the span of the set of distinct principal null directions has dimension four (maximally spanning type I) or dimension three (nonmaximally spanning type I). Explicit examples  are provided for both vacuum and nonvacuum spacetimes.
\end{abstract}

\keywords{Petrov type, principal null vectors}

\section{Introduction}

The Petrov classification of a spacetime Weyl curvature tensor is a local algebraic characterization based on the number of its distinct principal null directions (PNDs), represented by a set of at most four distinct null vectors modulo irrelevant rescaling factors. The corresponding sets of distinct principal null vectors are automatically linearly independent for all types except the Petrov type I case of four distinct such vectors where their span may have  either dimension 3 or 4 leading to a division of such cases into maximally spanning (dimension 4) or non-maximally spanning (dimension 3) type I cases.
The present refinement of the Petrov type I case based on the dimensionality of the span of a set of distinct principal null directions relies on a few basic properties of null vectors in a 4-dimensional Lorentzian spacetime. We use the word ``distinct" to describe a set of (nonzero) vectors such that no two of the vectors are proportional. Since the principal null directions determine bivector eigentensors of the curvature tensor, their overall scale is unimportant as with ordinary eigenvectors.

Since a 2-dimensional subspace contains at most 2 distinct null vectors, a third null vector which is not proportional to either one (and hence distinct) must be linearly independent so that the 3 vectors together determine a 3-dimensional Lorentzian subspace whose normal vector must be spacelike, i.e., a timelike subspace. These 3 null vectors then belong to the 2-dimensional light cone in that timelike subspace. A fourth distinct null vector either belongs to that lower dimensional light cone, or is linearly independent of the first 3 vectors, leading to a 4-dimensional span of the entire set.

Since the known exact solutions of Einstein's equations are very special with high symmetry, the Petrov type of their Weyl curvature tensor does not depend on position so the spacetime itself is said to be of that Petrov type. For a Petrov type I spacetime with a minimally spanning set of principal null vector fields, a unique spacelike unit vector field (modulo sign) exists which determines the orientation of their 3-dimensional spanning subspace within each tangent space. 

The Petrov classification \cite{Stephani:2003tm} categorizes Weyl curvature tensors by the generic number of distinct PNDs, which in turn translates into possible multiplicities of the roots of an eigenvalue problem involving bivectors which can have either simple (unrepeated) eigenvalues or repeated eigenvalues. The multiplicity types for the number of PNDs are
\begin{itemize}
\item[] Type I: four simple (four distinct),
\item[] Type II: one double and two simple (three distinct),
\item[] Type D: two double (two distinct),
\item[] Type III: one triple and one simple (two distinct),
\item[] Type N: one quadruple (one distinct),
\item[] Type O: none (vanishing Weyl tensor).
\end{itemize}
Type I is the algebraically general case, while the remaining types are referred to as algebraically special.

When studying  the  properties of a given spacetime, useful geometrical and physical information is associated with the principal null directions of its Weyl tensor.
Why do the PNDs play such an important role? A ``rough" argument is the following. The PNDs locate on the light cone at each spacetime point the pillars on which the spacetime itself can stand alone as a solution of the vacuum Einstein equations. If the spacetime then hosts other fields (either test fields or by generalization through perturbation fields which modify the background geometry through back-reaction), it is expected that the characteristic directions of these new fields will coincide, at least in a first approximation,  with those of the background.
This is true for the Petrov type D Kerr-Newman rotating and charged black hole spacetime, sourced by the electromagnetic field generated by a single massive electric charge. This spacetime generalizes the electrically neutral rotating Kerr black hole: the eigenvectors  of the electromagnetic field 2-form are aligned with those of the spacetime curvature. This agreement is captured by the vanishing of the generalized Simon tensor \cite{Bini:2004qf}.

Tr\"umper was the first to note the two possible spanning dimensions of a set of PNDs for a Petrov type I Weyl tensor \cite{Trumper} later mentioned in
 a general spinor discussion  pioneered by Rindler and Penrose \cite{Penrose:1986ca} and later studied by McIntosh et al \cite{ArMcIn1994}, who used invariants of the curvature tensor to give a condition for when this spanning dimension is not maximal, and in particular  
that if the Weyl tensor is either purely electric or purely magnetic (and therefore of type I), the span of the PNDs is only 3-dimensional. However, these conditions are not directly related to this dimensionality. Here we evaluate the wedge product of the 4 distinct PNDs to establish a direct connection between the PNDs and this dimensionality.

Analytically computing the PNDs of a given spacetime is always possible in principle,  but the actual computation can be quite difficult since it involves the roots of a fourth degree polynomial and their use in the subsequent bivector manipulations.
The usual approach starts with a null frame which is then conveniently ``rotated" (Lorentz transformed) until one of the frame vectors becomes a PND. In this case, spacetime symmetries may help, in the sense that a null vector $k_\pm$ is proportional to the sum or difference of a unit timelike vector $u$ and a unit spacelike one $\hat \nu$ orthogonal to $u$, $k_\pm \propto u \pm \hat \nu$, where either of these might be suggested by some Killing symmetries of the spacetime which might exist.

If one is interested only in characterizing the Petrov type of a given spacetime, it is enough to study the multiplicity of the PNDs without explicitly determining them.
However,
1) a dynamical spacetime (including perturbed black hole spacetimes and numerically generated spacetimes), during its evolution, may pass through different Petrov types, and it is interesting to study the motivations for this transition; 
2)
a general family of spacetimes, with a metric depending on several parameters, can also be of different Petrov types corresponding to various regions of the parameter space. Since different Petrov types are associated with distinct physical properties, it is interesting to study situations in which such changes happen. 

In particular distinguishing algebraically special spacetimes from the general type I case can be done by evaluating the ``speciality index" $\mathcal S$ \cite{Baker:2000zm,Beetle:2002iu}, 
a particular combination of the Weyl curvature scalars.
It has the value $\mathcal S=1$ only for algebraically special spacetimes, while $\mathcal S\neq 1$ characterizes the general type I case, thus identifying the algebraically special cases among a family of spacetimes which is generically of type I. 
This it true of the 1-parameter family of Kasner spacetimes which although generically of Petrov type I, allows isolated Petrov type D or O cases where  additional local rotational symmetry occurs; however, all type I Kasner spacetimes are found to be nonmaximally spanning. The Petrov exact sloution spacetime provides another explicit example of a Petrov type I spacetime, but which is maximally spanning.

Consider a Petrov type I  spacetime for which the 4 distinct PNDs are represented by the null vectors $k_i$, $i=1,\ldots 4$. These may span the entire tangent space or a 3-dimensional subspace, in which case their wedge product $\Omega_{1234}=k_1\wedge k_2\wedge k_3\wedge k_4$ is either nonzero (maximally spanning) or zero (nonmaximally spanning). The actual value when nonzero has no intrinsic meaning since the null vectors can be arbitrarily rescaled and only their equivalence classes under rescaling matter (algebraically).
In contrast for the case of the Petrov type II spacetimes, where there exist only three distinct PNDs $k_i$, $i=1,\ldots 3$, their wedge product $k_1\wedge k_2\wedge k_3$ is automatically nonzero as discussed above (and its dual defines a spacelike normal to their span),  so this dimensional distinction is no longer relevant for it or the remaining Petrov types in the hierarchy.

Apart from the clear geometrical meaning of this division of Petrov type I cases in terms of the spanning set dimension, its physical meaning is not yet apparent and will require further investigation. For example, what role does the spacelike normal to the spanning set play in the nonmaximal type I case? In the Kasner case it turns out to be associated spatial direction with the single negative Kasner index leading to contraction in the forward time direction.

Our conventions and notation will follow
the standard ones for the Newman-Penrose (NP) 
formalism~\cite{Teukolsky:1973ha,Chandrasekhar:1985kt} (see also Ref.~\cite{Stephani:2003tm}).
Furthermore, units are chosen such that $c=1=G$ and the metric signature is $+---$ as usually chosen when using the NP formalism.

\section{Petrov classification and scalar invariants: a short review}

Consider the Weyl tensor $C_{\alpha\beta\gamma\delta}$ of a given spacetime with metric $g$ and its dual ${}^*C_{\alpha\beta\gamma\delta}$.
Define the complex tensor 
$\tilde C_{\alpha\beta\gamma\delta}=C_{\alpha\beta\gamma\delta}-i{}^*C_{\alpha\beta\gamma\delta}$
and introduce in both tensor and Newman-Penrose (NP) notation the two
complex curvature invariants \cite{Stephani:2003tm}
\begin{equation}
\label{Idef}
I=\frac 1{32}\tilde{C}_{\alpha\beta\gamma\delta}\tilde{C}^{\alpha\beta\gamma\delta}
=\psi _0\psi _4-4\psi _1\psi_3+3\psi _2^2\,,
\end{equation}
and 
\begin{eqnarray}
\label{Jdef}
J&=&\frac 1{384}\tilde{C}_{\alpha\beta\gamma\delta}\tilde{C}^{\gamma\delta}{}_{\mu\nu}\tilde{C}^{\mu\nu\alpha\beta}
=\psi_0\psi _2\psi _4-\psi _1^2\psi _4-\psi _0\psi _3^2+2\psi _1\psi _2\psi_3-\psi _2^3\,,
\end{eqnarray}
where the Weyl scalars refer to a choice of NP frame $\{l, n, m, \bar m\}$ related to an associated orthonormal frame $\{e_\alpha\}=\{e_0,e_a\}$ by the standard relations
\beq
\label{NPvsorthon}
l=\frac{1}{\sqrt{2}}(e_0+e_1)\,, \qquad
n=\frac{1}{\sqrt{2}}(e_0-e_1)\,, \qquad
m=\frac{1}{\sqrt{2}}(e_2+i e_3) \,.
\eeq

An observer with 4-velocity $U$ measures the following electric and magnetic parts of the Weyl tensor
\beq
E(U)_{\alpha\beta}=C_{\alpha\mu\beta\nu}U^\mu U^\nu, \quad  
H(U)_{\alpha\beta}=-{}^*C_{\alpha\mu\beta\nu}U^\mu U^\nu \,,
\eeq
respectively, which can be combined into the symmetric tracefree complex tensor 
\beq
Q(U)_{\alpha\beta}=\tilde C_{\alpha\mu\beta\nu}U^\mu U^\nu
=E(U)_{\alpha\beta}+iH(U)_{\alpha\beta}\,,
\eeq
in terms of which the scalars $I$ and $J$ take the form
\bea
I&=&\frac 1{32}Q(U)^\alpha{}_\beta Q(U)^\beta{}_\alpha\,,
\quad 
J=\frac 1{384}Q(U)^\alpha{}_\beta Q(U)^\beta{}_\delta Q(U)^\delta{}_\alpha\,.
\eea
Let $e_0=U$, so that the orthonormal frame $\{e_\alpha\}$ is adapted to the observer $U$.
The nonzero components of the tensor $Q$ with respect to it can be represented by the following $3\times3$ complex matrix 
\beq
(Q^a{}_b)=\left(
\begin{array}{ccc}
\psi_2-\frac12 (\psi_0+\psi_4)&\frac{i}{2}(\psi_4-\psi_0)&\psi_1-\psi_3\cr
\frac{i}{2}(\psi_4-\psi_0)&\psi_2 +\frac12 (\psi_0+\psi_4)&i(\psi_1+\psi_3)\cr
\psi_1-\psi_3&i(\psi_1+\psi_3)&-2\psi_2 \cr
\end{array}
\right)
\,,
\eeq
where $a,b=1,2,3$.

The scalars $I$ and $J$  are used to define the speciality index $\mathcal{S}$~\cite{Baker:2000zm,Beetle:2002iu} of the spacetime when $I\neq0$
\begin{equation}
\label{SPECT}
\mathcal{S}=\frac{27J^2}{I^3}\,,  
\end{equation}
characterizing the transition from general Petrov type $I$ ($\mathcal{S}\neq 1$) to algebraically special behavior ($\mathcal{S}=1$)~\cite{Stephani:2003tm}.
Because of their tensor expressions as scalars, it is clear that both $I$ and $J$ (and hence $\mathcal{S}$) are frame-invariant objects, i.e., they do not change under any allowed transformation of the chosen orthonormal or null frame.

The standard algorithm used to determine the Petrov type of a given spacetime involves the evaluation of other scalar objects. 
One first evaluates the scalars $I$ and $J$ and the difference $I^3-27J^2$. If the latter quantity is nonzero then the spacetime is of type I. If instead $I^3-27J^2=0$ one should distinguish the case of $I$ and $J$ both nonvanishing or not, and construct three new scalars \cite{Stephani:2003tm},
\bea
\label{scalars}
K&=&\psi_1\psi_4^2-3\psi_4\psi_3\psi_2+2\psi_3^3\,, \nonumber\\
L&=&\psi_2\psi_4-\psi_3^2\,, \nonumber\\
N&=&12L^2-\psi_4^2 I\,,
\eea 
which are related to the discriminants of the quartic equation \eqref{lambda_eq} defining the PNDs, and are not frame-invariant (see Appendix A).
The algebraically special types correspond to the following conditions:
\begin{eqnarray}
&&\hskip-20pt\mbox{Type II:\quad} I\not=0\,,\ J\not =0\,,\ K\not=0\,\,\ {\rm or/and}\,\,\ N\not =0 \,,\nonumber
\\
&&\hskip-20pt\mbox{Type D:\quad} I\not=0\,,\  J\not =0\,,\  K=0\,,\  N =0 \,,\nonumber
\\
&&\hskip-20pt\mbox{Type III:\ } I=0\,,\  J =0\,,\  L\not=0\,\,\ {\rm or/and}\,\,\ K\not =0 \,,\nonumber
\\
&&\hskip-20pt\mbox{Type N:\ } I=0\,,\  J =0\,,\  L=0\,,\  K=0 \,.
\end{eqnarray}
Details of this algorithm as well as its representation as a  flow chart can be found in Fig.~9.1 of Ref.~\cite{Stephani:2003tm}, 
recalling the underlying assumption   $\psi_4\not =0$  (or $\psi_0\not =0$).

An equivalent approach to classifying the Weyl tensor instead solves the eigenvalue problem associated with the matrix $Q_{ab}$.
The matrix criteria for the various Petrov types and the normal forms of the matrix $Q_{ab}$ in each case (with corresponding eigenvalues and eigenvectors) are listed in Tables 4.1 and 4.2 of 
Ref.~\cite{Stephani:2003tm}, respectively. 
The orthonormal frame $\{e_\alpha\}$ with respect to which the matrix $Q_{ab}$ has a normal form is uniquely determined (modulo the choice of numbering of the three spatial vectors $\{e_a\}$) for the non-degenerate Petrov types I, II and III, and is called a Weyl principal (or canonical) tetrad. 
The eigenvalues satisfy the equation
\beq
\sigma^3-I\sigma-2J=0\,,
\eeq
so that 
\bea
I&=&\frac12(\sigma_1^2+\sigma_2^2+\sigma_3^2)\,,
\quad 
J=\frac16(\sigma_1^3+\sigma_2^3+\sigma_3^3)=\frac12\sigma_1\sigma_2\sigma_3\,.
\eea
For Petrov type I spacetimes the Weyl scalars with respect to the principal tetrad are given by $\psi_0=\psi_4=(\sigma_2-\sigma_1)/2$, $\psi_1=\psi_3=0$, and $\psi_2=-\sigma_3/2$, with $\sigma_3=-\sigma_1-\sigma_2$.
For type D we have in addition $\psi_0=0=\psi_4$ as $\sigma_1=\sigma_2$.
For type II we have $\psi_0=\psi_1=\psi_3=0$, $\psi_4=-2$, and $\psi_2=-\sigma_3/2$.

\subsection{PNDs for Petrov types I and II spacetimes}

Next we review the explicit determination of the PNDs.
Following the notation of Ref.~\cite{Stephani:2003tm},  if $\psi_4\not =0$ for the Petrov type of a given spacetime we have to find the roots $\lambda$ (with the corresponding multiplicity) of the following algebraic equation 
\beq
\label{lambda_eq}
\lambda^4 \psi_4-4\lambda^3\psi_3 +6 \lambda^2\psi_2-4\lambda \psi_1+\psi_0=0\,,
\eeq
whose solutions define the explicit expressions for the four PNDs 
\beq
\label{pnds_general}
k_{i}=l+\lambda_i^*m +\lambda_i \bar m+|\lambda_i|^2 n\,,\quad i=1\ldots 4\,.
\eeq

These roots are computed as follows.
First divide Eq.~\eqref{lambda_eq} through by its leading coefficient
\beq
\label{lambda_eq2}
\lambda^4 +a_1 \lambda^3 +a_2 \lambda^2+a_3\lambda +a_4=0\,,
\eeq
defining the new coefficients
\beq
\label{aidef}
a_1=-\frac{4\psi_3}{\psi_4}\,,\quad a_2=\frac{6\psi_2}{\psi_4}\,,\quad a_3=-\frac{4\psi_1}{\psi_4}\,,\quad a_4=\frac{\psi_0}{\psi_4} \,.
\eeq
This equation can be directly solved by using the standard, rather involved, formulas available in the literature, leading to the four roots $\lambda_i$ ($i=1,\ldots 4$).
However, one can conveniently rotate the NP frame to put it into its transverse form, i.e., with $\psi_1=0=\psi_3$, so that Eq.~\eqref{lambda_eq2} reduces to a bi-quadratic equation
\beq
\lambda^4 +a_2 \lambda^2 +a_4=0\,,
\eeq
with solutions
\beq
\label{lambdas_1234}
\lambda_{1,2}=\Lambda_\pm\,,\qquad \lambda_{3,4}=-\Lambda_\pm  \,, 
\eeq
where
\beq
\Lambda_\pm =\sqrt{\frac{-a_2\pm \sqrt{a_2^2-4a_4}}{2}}\,.
\eeq
If the transverse frame is also canonical ($\psi_0=\psi_4$, $a_4=1$), additional simplifications in the solutions \eqref{lambdas_1234} occur, namely
\beq
\Lambda_\pm =\frac{1}{2}\left[\sqrt{ -a_2+2}\pm \sqrt{-a_2-2}\right]\,.
\eeq
For example, for Petrov type I, inserting the value of $a_2={6 \psi_2}/{\psi_0}$ 
leads to the following (explicit) solutions \cite{ArMcIn1994} 
\beq
\label{lambdasIcanframe}
\lambda_1\,, \quad 
\lambda_2=-\lambda_1\,, \quad
\lambda_3=\frac1{\lambda_1}\,, \quad
\lambda_4=-\frac1{\lambda_1}\,,
\eeq 
where
\beq
\label{lambda1sol}
\lambda_1=\left[-3\frac{\psi_2}{\psi_0}-\sqrt{9\left(\frac{\psi_2}{\psi_0}\right)^2-1}\right]^{1/2}\,.
\eeq
The latter can be also written as 
\beq
\label{lambda1soln}
\lambda_1=\frac{\sqrt{\sigma_2+2\sigma_1}+ \sqrt{\sigma_1+2\sigma_2}}{\sqrt{\sigma_1-\sigma_2}}\,,
\eeq
in terms of the eigenvalues $\sigma_i$ of the matrix $Q_{ab}$ (see Table 4.3 of Ref.~\cite{Stephani:2003tm}).

On the one hand, directly solving the fourth-degree algebraic equation \eqref{lambda_eq} is in general a difficult task (because of the large expressions involved), but which is  facilitated if one uses a principal NP frame. In fact, in that case this equation becomes bi-quadratic with obvious advantages in writing its solutions.

On the other hand, transforming a general NP frame into a principal one is not an easy task, since one generally must use type I, II and III null tetrad rotations in succession to accomplish this, a fact which in most cases works against the advantage of solving a simpler equation at the end.

In the case of Petrov type II spacetimes the canonical tetrad corresponds to $\psi_0=0=\psi_1=\psi_3$ and $\psi_4=-2$ \cite{Stephani:2003tm}, so that Eq.~\eqref{lambda_eq} becomes
\beq
\label{lambda_eq_II}
-2\lambda^2(\lambda^2-3\psi_2)=0\,,
\eeq
with solutions 
\beq
\label{lambdasIIcanframe}
\lambda_1=0=\lambda_2\,, \quad 
\lambda_3=\sqrt{3\psi_2}\,, \quad
\lambda_4=-\lambda_3 \qquad (\psi_2\neq0)
\,.
\eeq 
Therefore, $k_1=l=k_2$ is a repeated PND with multiplicity 2, while $k_3,k_4$ are given by Eq.~\eqref{pnds_general}.
On the other hand
the complex matrix $Q_{ab}$ has eigenvalues $\sigma_1=\sigma_2=-\sigma/2$ and $\sigma_3=\sigma=-2\psi_2$, so that $\lambda_3=\sqrt{-\frac32\sigma}$.

\section{PND degeneracy: a geometrical approach}

The four PNDs \eqref{pnds_general} may be either linearly independent or not.
In the former case they span a 4-dimensional vector space at each spacetime point, otherwise only a 3-dimensional subspace.

Arianrhod, McIntosh and coworkers \cite{ArMcIn1994,ArMcIn1990,ArMcIn1992} classified the PND degeneracies depending on the nature and value of the scalar invariant 
\beq
\label{tildeMdef}
\tilde M=\frac{I^3}{J^2}-27=\frac{27}{\mathcal S}(1-\mathcal S)\,,
\eeq
with $\tilde M$ generally complex and possibly infinite.%
\footnote{
We denote here such an invariant by $\tilde M$ instead of $M={I^3}/({J^2}-6)$ since we are using the definitions of Ref.~\cite{Stephani:2003tm} for $I$ and $J$, which slightly differ from those of Penrose and Rindler \cite{Penrose:1986ca}.
}
They proved the following theorem \cite{ArMcIn1990}: ``The four distinct PNDs associated with a metric whose Weyl tensor is of Petrov type I span, at each point, either a 3-dimensional vector space, in which case $\tilde M$ is real and either positive or infinite, or a 4-dimensional vector space for other $\tilde M$.''
Furthermore, they showed that if there exists an observer with 4-velocity $U$ who sees the Weyl tensor as purely electric or purely magnetic, then the PNDs are linearly dependent, and span the 3-dimensional vector space orthogonal to the eigenvector of $Q_{ab}$ corresponding to the eigenvalue of smallest absolute value \cite{ArMcIn1994}.

Here we will adopt a different criterion (leading, however, to equivalent conclusions) to distinguish between the two cases.
Let 
\beq
\Omega_{1234}=k_1\wedge k_2\wedge k_3\wedge k_4
\eeq
be the 4-dimensional volume associated with the $k_i$.
When $\Omega_{1234}\not=0$ the four PNDs are linearly independent, our maximally spanning type I case.
The nonmaximally spanning type I case instead corresponds to $\Omega_{1234}=0$, implying that the PNDs are linearly dependent.

In general, the volume 4-form has the expression
\beq
\label{4volpnd}
\Omega_{1234}={\mathcal V}\, l\wedge n \wedge m\wedge \bar m\,,
\eeq
with
\bea
\label{calVdef}
{\mathcal V}&=&(\lambda_{32}+\lambda_{24}-\lambda_{34})|\lambda_1|^2 +(\lambda_{13}+\lambda_{34}-\lambda_{14})|\lambda_2|^2\nonumber\\
&+& (\lambda_{14}+\lambda_{21}-\lambda_{24})|\lambda_3|^2 +(\lambda_{12}+\lambda_{23}-\lambda_{13})|\lambda_4|^2 \,,
\eea
where we have used the notation
\beq
\lambda_{nm}=\bar \lambda_n \lambda_m -\lambda_n \bar \lambda_m=\bar \lambda_{mn}\,.
\eeq
The quantity ${\mathcal V}$ vanishes identically when all the $\lambda_i$ are either real or purely imaginary, which leads to $\lambda_{mn}=0$ for all $m,n$, or when all the $\lambda_i$ are unit complex numbers $|\lambda_i|=1$, when the expression reduces to
\beq
 {\mathcal V} = (\lambda_{12}+\lambda_{21}) + (\lambda_{23}+\lambda_{32})
\eeq
which vanishes since in this case $\lambda_{nm}=-\lambda_{mn}$.
${\mathcal V}$ may be different from zero only if the (distinct) $\lambda_i$ are non-unit complex numbers, which is a necessary and sufficient condition for linear independence.

For Petrov type I spacetimes, substituting into Eq.~\eqref{calVdef} the solutions
 \eqref{lambdasIcanframe} corresponding to a canonical tetrad leads to the expression
\beq\label{Vee}
{\mathcal V}
=-\frac{16}{|\lambda_1|^4}{\rm Re}(\lambda_1){\rm Im}(\lambda_1)(|\lambda_1|^2+1)(|\lambda_1|^2-1)\,,
\eeq
implying that the PNDs are linearly dependent only if one of the following conditions holds: 
${\rm Re}(\lambda_1)=0$, ${\rm Im}(\lambda_1)=0$, or  $|\lambda_1|^2=1$. These are the same conditions on all the eigenvalues which holds in general, but for the type I case the interrelationships of these eigenvalues makes it sufficient to hold only for one of them to hold for all of them.

For Petrov type II spacetimes, the 3-dimensional volume associated with the canonical tetrad reads
\bea
\label{Omega123can}
\Omega_{123}&=&k_1 \wedge k_2 \wedge k_3\nonumber\\
&=&-2|\lambda_3|^2\,l\wedge n\wedge \left(\bar\lambda_3 m+\lambda_3\bar m\right)\nonumber\\
&=&2\sqrt{2}|\lambda_3|^2\left[{\rm Re}(\lambda_3)\,\omega^{012}+{\rm Im}(\lambda_3)\,\omega^{013}\right]
\,,
\eea
with $\omega^{012}=\omega^0\wedge \omega^1\wedge \omega^2$ and $\omega^{013}=\omega^0\wedge \omega^1\wedge \omega^3$, where  $\{\omega^\alpha\}$ is the dual frame of $\{e_\alpha\}$, related to the NP frame in the usual way by Eq.~\eqref{NPvsorthon}, while $\lambda_3=\sqrt{3\psi_2}\neq0$ cannot vanish and remain of type II.
Therefore, it is always nonzero, implying that as expected, there cannot exist spacetimes of nonmaximally spanning type II. 
The nonexistence of type II spacetimes with linearly dependent PNDs has not been pointed out before, and is a novel and unexpected result of the present analysis.

In Section 5 we will consider explicit examples which prove helpful by illustrating the previous discussion concretely.

\section{Relation with the algebraic approach of Arianrhod and McIntosh}

We now show for Petrov type I the equivalence  between our geometrical approach (based on the vanishing of the 4-dimensional volume element \eqref{4volpnd} associated with the PNDs, or the scalar quantity ${\mathcal V}$, Eq.~\eqref{Vee}) and the algebraic criterion of Arianrhod and McIntosh \cite{ArMcIn1990,ArMcIn1992} referred to in the previous section (based on the value of the scalar invariant $\tilde M$, Eq.~\eqref{tildeMdef}).
The latter can be expressed in terms of the canonical tetrad as follows
\beq
\tilde M=\frac{2916(\lambda_1^4-1)^4\lambda_1^4}{(1+\lambda_1^4)^2(\lambda_1^4+6\lambda_1^2+1)^2[(\lambda_1^2-1)^2-4\lambda_1^2]^2}\,,
\eeq
with $\lambda_1\not=0$ given by Eq.~\eqref{lambdasIcanframe}. Unfortunately this brute force proof requires a computer algebra system to accomplish because of the complicated relationship between $\tilde M$ and $\lambda_1$. 

Linear dependence of the PNDs requires that the imaginary part of $\tilde M$ vanish, while the real part must be positive or infinite, so it must be shown that this is equivalent to the vanishing of $\mathcal V$.
Introduce the real and imaginary parts of $\lambda_1=a+ib$, in terms of which $\mathcal V$ is the explicitly real expression for type I spacetimes
\beq
{\mathcal V}
=\frac{16ab\,[1-(a^2+b^2)^2]}{(a^2+b^2)^2}\,.
\eeq

Unfortunately the following brute force proof requires a computer algebra system to accomplish because of the complicated relationship between $\tilde M$ and $\lambda_1$. 
Introducing some auxiliary complex quantities $x+iy,z+iw$ defined below leads to the following expression for $\tilde M$
\beq
\tilde M=\frac{2916\,(x+iy)^4}{(z+iw)^2}\,,
\eeq
and hence 
\bea
{\rm Re}(\tilde M)&=&-\frac{2916}{(z^2+w^2)^2}[(x^2-y^2)(w+z)+2xy(w-z)]\nonumber\\
&&
\times [(x^2-y^2)(w-z)-2xy(w+z)]
\,,\nonumber\\
{\rm Im}(\tilde M)&=&-\frac{5832}{(z^2+w^2)^2}[(x^2-y^2)w-2xyz][(x^2-y^2)z+2xyw]
\,.
\eea
The quantities $x$, $y$, $z$, $w$ are defined by ugly expressions
\bea
x&=&a\,(a^4-10a^2b^2+5b^4-1)
\,,\nonumber\\
y&=&b\,(b^4-10a^2b^2+5a^4-1)
\,,\nonumber\\
z&=&1+198a^2b^2-33b^4-33a^4+924a^6b^2-2310a^4b^4+924a^2b^6-66a^{10}b^2\nonumber\\
&&
+495a^8b^4-924a^6b^6+495a^4b^8-66a^2b^{10}-33a^8-33b^8+a^{12}+b^{12}
\,,\nonumber\\
w&=&4ab\,(a^2-b^2)(3a^8-52a^6b^2-66a^4+146a^4b^4-52a^2b^6+396a^2b^2\nonumber\\
&&
-33+3b^8-66b^4)
\,.
\eea
The imaginary part of $\tilde M$ vanishes if either
1) $(x^2-y^2)w-2xyz=0$, implying ${\rm Re}(\tilde M)=11664x^2y^2/w^2\geq0$, or if 
2) $(x^2-y^2)z+2xyw=0$, implying ${\rm Re}(\tilde M)=-2916(x^2-y^2)^2/w^2\leq0$.
The second case does not lead to linear dependence.
The first case instead leads to
\beq
0=2ab\,[(a^2+b^2)^2-1]P(a,b)\,,
\eeq
where $P(a,b)=P(b,a)\ge1$ is a symmetric real positive polynomial function which never vanishes.
Therefore, both conditions ${\rm Im}(\tilde M)=0$ and ${\rm Re}(\tilde M)\geq0$ are satisfied if and only  either $a=0$ or $b=0$ or $a^2+b^2=1$, implying that ${\mathcal V}=0$ as well. For completeness we note that
\beq
P(a,b)=Q(a,b) Q(-a,b) Q(a,-b) Q(-a,-b)\,,
\eeq
where
\begin{eqnarray}
Q(a,b)=(a^2+b^2+1)^2+2(a+b-1)(a^2 + b^2 + a + b)\,.
\end{eqnarray}

Consider the converse situation where
${\mathcal V}$ vanishes. This occurs in the three cases $a=0$ or $b=0$ or $a\neq0 \neq b, a^2+b^2=1$.
We have already shown that in every such case the imaginary part of $\tilde M$ is identically vanishing. 
Concerning the real part, if $a=0$ it reduces to 
\beq
{\rm Re}(\tilde M)=\frac{2916(b^4-1)^4b^4}{(1+b^4)^2(b^4+6b^2+1)^2[(b^2-1)^2-4b^2]^2}\,,
\eeq
while the case $b=0$ is equivalent to this exchanging $a$ and $b$, while if $a^2+b^2=1$, we can use $a^2=1-b^2$ to re-express this quantity as
\beq
{\rm Re}(\tilde M)=\frac{729(b^2-1)^2b^4}{(b^2-2)^2(2b^2-1)^2(1+b^2)^2}\,.
\eeq
In all three cases either $\tilde M\geq0$ or $\tilde M$ is infinite.
Therefore, ${\mathcal V}=0$ implies ${\rm Im}(\tilde M)=0$ and ${\rm Re}(\tilde M)\geq0$.

However, the quantity $\tilde M$ is an unmotivated combination of the two complex curvature scalars associated with the algebraic classification of the curvature tensor and the properties of this quantity which lead to linear dependence of the PNDs are awkward and without direct interpretation. In contrast the quantity $\mathcal V$ is directly associated with the volume form determined by the PNDs, with an immediate interpretation of its vanishing or nonvanishing in terms of the linear independence of the PNDs.

\section{Type I spacetimes: examples}

\subsection{Kasner spacetime}

The simplest Petrov type I spacetime allowing for analytical computations is the vacuum Kasner~\cite{Landau:1975pou} metric 
\begin{equation}
ds^2=dt^2-t^{2p_1}d x^2-t^{2p_2}dy^2-t^{2p_3}dz^2\,,  \label{LLLL}
\end{equation}
where the so-called Kasner indices $p_i$ satisfy
\begin{equation}
p_1+p_2+p_3=p_1^2+p_2^2+p_3^2=1\,, \label{constr}
\end{equation}
and assume values in the closed interval $[-\frac13,1]$.
The spatial Cartesian coordinates and these indices are adapted to the eigenvectors of the extrinsic curvature of the intrinsically flat time slices, but the algebraic properties of the spacetime curvature tensor are quite different. 
 
Introduce the following NP frame adapted to the first spatial coordinate
\begin{eqnarray}  
\label{tetrade}
l &=& \frac{1}{\sqrt{2}}[\partial_t+t^{-p_1}\partial_x], \nonumber\\
  n&=&\frac{1}{\sqrt{2}}[\partial_t-t^{-p_1}\partial_x] \,,  \nonumber\\
m &=&\frac 1{\sqrt{2}}[t^{-p_2}\partial_y+it^{-p_3}\partial_z] \,,  
\end{eqnarray}
which has the following nonzero Weyl scalars 
\begin{eqnarray}
\label{weylscal}
&&\psi _0=\psi_4=\frac{p_1(p_2-p_3)}{2t^2}, \quad \psi _2=-\frac{p_2p_3}{2t^2}\,,
\end{eqnarray}
so that the frame is a canonical one (clearly true starting from the other two coordinate directions as well).
The associated orthonormal frame 
\beq\label{orto}
e_0=\partial_t\,,\quad
e_1=t^{-p_1}\partial_x\,,\quad
e_2=t^{-p_2}\partial_y\,,\quad
e_3=t^{-p_3}\partial_z\,,
\eeq
is adapted to the static observers with 4-velocity $U=e_0$ whose spatial axes are aligned with the Killing vectors $\partial_x, \partial_y, \partial_z$, and therefore directly observe the homogeneity of the spacetime. They also see a purely electric Weyl tensor whose electric part is
\beq
\label{elemagn}
E(U)=\frac{1}{t^2}\left[ p_1p_3 \, e_1\otimes e_1+  p_1p_2 \, e_2\otimes e_2+  p_2p_3 \, e_3\otimes e_3\right]\,, 
\eeq
while its magnetic part $H(U)$ vanishes identically.

The 1-parameter family of spacetimes \eqref{LLLL} is efficiently parametrized by expressing the Kasner indices in terms of the Lifshitz-Khalatnikov (LK) parameter
\beq
\label{LK_param}
p_1 = -\frac{u}{(1+u+u^2)}\,,\qquad 
p_2 = \frac{(1+u)}{(1+u+u^2)}\,,\qquad 
p_3 = \frac{u (1+u)}{(1+u+u^2)}\,,
\eeq
with limiting cases $u\rightarrow \pm\infty$ capturing the remaining triplet $(0,0,1)$, while $u=0\leftrightarrow (0,1,0)$ and $u=-1\leftrightarrow (1,0,0)$, all of which correspond to a flat spacetime.
On the other hand the three cases $u=-2,-\frac12,1$ correspond to the three triplets which are permutations of $(-\frac13,\frac23,\frac23)$ for which the spacetime is a locally rotationally symmetric type $D$ spacetime, with a spindle-like cosmological singularity~\cite{Stephani:2003tm,DIN}, expanding in one direction while collapsing in the two orthogonal directions. 

The parameter space of these Kasner spacetimes is best understood as a circle with three equal divisions separated by the three flat spacetime values, and with the locally rotationally symmetric cases at the center of each such interval. See Fig.~2 of \cite{bini2007}.
These three intervals are those for which one Kasner index is negative and the other two positive, corresponding to contraction in that direction as the spacetime evolves: 
$p_1<0: -1<u^{-1}<1$,
$p_2<0: -\infty<u<-1$,
$p_3<0:-1<u<0$.

Using the LK parametrization, the various scalars turn out to be
\bea
I&=& \frac{u^2 (1+u)^2}{  (1+u+u^2)^3} \frac1{t^4}\,,\qquad
J= \frac12 \frac{u^4(1+u)^4 }{(1+u+u^2)^6 } \frac1{t^6}\,, \nonumber\\
K&=& 0\,, \qquad
L=  -\frac14 \frac{u^2 (u-1) (1+u)^3 }{(1+u+u^2)^4 } \frac1{t^4}\,, \nonumber\\
N&=& \frac{ u^4(u+2) (2 u+1) (u-1)^2 (1+u)^4}{4 (1+u+u^2)^8 } \frac1{t^8}\,.
\eea
Note that both $I$ and $J$ vanish for the flat cases $u=0,-1,\pm\infty$ but the speciality index is nevertheless always defined and has the constant value
\beq
\mathcal{S} = -\frac{27}{4}p_1p_2p_3
= \frac{27}{4}\frac{u^2 (1+u)^2}{(1+u+u^2)^3}
\,,  \label{SPECTRX}
\eeq
where these expressions here (and their permutations) are equivalent due to (\ref{constr}).
Apart from $K$ which vanishes identically, other zero values of the remaining NP scalars do occur. 
All of these scalars vanish for the trivial flat spacetime case for which $u=0,-1,\pm\infty$, while for the three type D cases where $u=-2,-\frac12,1$ one has
$I=\frac{4}{27}t^{-4}$, $J=\frac{8}{729}t^{-6}$ and $u=1$: $L=N=0$, $u=-2,-\frac12$: $N=0$. 
Thus within the Kasner family transitions of Petrov type only occur among types I, D, and O at these particular parameter values.

Consider now the PNDs in the type I case.
The Arianrhod-McIntosh invariant \eqref{tildeMdef} reads
\beq
\tilde M=\frac{ (u+2)^2(2 u+1)^2(u-1)^2}{u^2 (1+u)^2}\,,
\eeq
and it is always positive for every value of $u\neq1$ and therefore real, implying that the four PNDs must be linearly dependent for all finite values of $u$ except the trivial case $u=1$ of flat spacetime.

On the other hand we can derive this result directly. The frame \eqref{orto} is a canonical frame, so that the PNDs are given by Eq.~\eqref{pnds_general} with $\lambda_i$ specified by Eq.~\eqref{lambdasIcanframe}.
The eigenvalues of the matrix $Q_{ab}$ are 
\beq
\sigma_1=u\,\sigma_2=-\frac{u^2(1+u)}{(1+u+u^2)^2}\frac1{t^2}\,,
\eeq
so that by Eq.~\eqref{lambda1soln}
\beq
\lambda_1=\sqrt{\frac{u+2}{u-1}}+\sqrt{\frac{2u+1}{u-1}}\,,
\eeq
which is real for $u>1$ and $u<-2$, purely imaginary for $-\frac12<u<1$, and complex for $-2<u<-\frac12$ with $|\lambda_1|^2=1$, all conditions for which by Eq.~\eqref{Vee}
lead to ${\mathcal V}=0$ implying linear dependence of the PNDs for all values of $u\neq1$.

Finally we can evaluate the three 3-vectors obtained by wedging together 
each triplet combination of the PND's, namely
\beq
\label{k_abc}
k_{234}\,,\quad 
k_{134}\,,\quad
k_{124}\,,\quad
k_{123}\,,
\eeq
with $k_{abc}=k_a\wedge k_b\wedge k_c$.
If $p_a<0$, we find that each of the $k_{abc}$ in Eq.~(\ref{k_abc}) is proportional 
 to  $\omega^0\wedge\omega^b\wedge\omega^c$, where $(a,b,c)$ is a cyclic 
permutation of (1,2,3). In other words, the span of the 4 PND's is the 
subspace orthogonal to the single collapsing spatial direction $e_a$. 

Apparently, the algebraic Petrov-type properties of the spacetime curvature reflect the distinction between collapsing and expanding spatial directions in its global 
light cone structure through the Einstein equations. 
This is a new observation that has escaped the notice of previous investigations
and sheds some light on physical consequences of the 
Petrov classification.

\subsection{Petrov spacetime}

The Petrov spacetime \cite{Petrov1962} is a homogeneous vacuum solution with line element given by
\beq
k^2 ds^2= e^x [\cos (\sqrt{3}x)(dt^2-dz^2)+2\sin (\sqrt{3}x) dt dz]-dx^2-e^{-2x}dy^2\,,
\eeq
where $k>0$ is a constant parameter and $0<\sqrt{3}x<\pi/2$.
The orthonormal frame associated with the principal NP frame is given by
\bea
e_0&=& ke^{-x/2}\left[\cos \left(\frac{\sqrt{3}x}{2}\right)\partial_t+\sin \left(\frac{\sqrt{3}x}{2}\right)\partial_z\right]
\,,\nonumber\\
e_1&=& \frac{k}{\sqrt{2}}(\partial_x-e^{x}\partial_y)
\,,\nonumber\\
e_2&=& \frac{k}{\sqrt{2}}(\partial_x+e^{x}\partial_y)
\,,\nonumber\\
e_3&=& ke^{-x/2}\left[-\sin \left(\frac{\sqrt{3}x}{2}\right)\partial_t+\cos \left(\frac{\sqrt{3}x}{2}\right)\partial_z\right]
\,,
\eea
leading to the following nonvanishing Weyl scalars
\bea
\psi_0&=&\psi_4=-\frac{k^2\sqrt{3}}{2}e^{i\pi/6}
\,,\nonumber\\
\psi_2&=& -\frac{k^2}{2}e^{-i\pi/3}=-k^2-\psi_4\,.
\eea
The various NP scalars are explicitly
\beq
I = 0\,,\quad
J = -\frac{k^6}{2}\,,\quad
K = 0\,,\quad
L = \frac{\sqrt{3}\,k^4}{4}e^{-i\pi/6}\,,\quad
N=\frac{9k^8}{4}e^{-i\pi/3}\,,
\eeq
so that the Arianrhod-McIntosh invariant \eqref{tildeMdef} is negative ($\tilde M=-27$).
Both the electric and magnetic parts of the Weyl tensor measured by $U=e_0$ are nonzero
\bea
E(U)&=&\frac{k^2}{2}\left(e_1\otimes e_1+e_2\otimes e_2- 2\, e_3\otimes e_3\right)
\,,\nonumber\\
H(U)&=&\frac{\sqrt{3}k^2}{2}\left(e_1\otimes e_1-e_2\otimes e_2\right)
\,.
\eea 

The complex matrix $Q_{ab}$ has eigenvalues
\beq
\sigma_1=-k^2e^{i\pi/3}\,,\quad 
\sigma_2=-k^2=e^{i\pi}\,,\quad
\sigma_3= k^2e^{-i\pi/3}\,,
\eeq
so that from Eq.~\eqref{lambda1soln}
\beq
\lambda_1=e^{-i\pi/3}+e^{-i\pi/6}
=\frac12(1-i)(1+\sqrt{3})\,,
\eeq
and $\mathcal V=16\sqrt{3}$, implying linear independence of the PNDs.

\subsection{Static spacetimes of the Weyl class}

Static axisymmetric vacuum solutions of the Einstein field equations can be described using Weyl's approach \cite{weyl}. The line element in cylindrical coordinates $(t,\rho,z,\phi)$ has the form
\begin{equation}
\label{weylmetric}
ds^2=e^{2\psi}dt^2-e^{2(\gamma-\psi)}(d\rho^2+dz^2)-\rho^2e^{-2\psi}d\phi^2\,,
\end{equation}
where the functions $\psi$ and $\gamma$ only depend on the coordinates $\rho$ and $z$. 
The vacuum Einstein field equations reduce to a decoupled second order equation (the axisymmetric Laplace equation in flat space) and two first order equations
\begin{eqnarray}
\label{einsteqs}
0&=&\psi_{,\rho\rho}+\frac1\rho\psi_{,\rho}+\psi_{,zz}
\,,\nonumber\\
0&=&\gamma_{,\rho}-\rho(\psi_{,\rho}^2-\psi_{,z}^2)
\,,\nonumber\\
0&=&\gamma_{,z}-2\rho\psi_{,\rho}\psi_{,z}\,.
\end{eqnarray}
The linearity of the  first equation allows explicit spacetime solutions representing superpositions of two or more axially symmetric bodies, which turn out to be Petrov type I. Other solutions for single axially symmetric bodies, however,  exist  and  are in general of Petrov type D. We will limit our considerations here to  the general case of the metric \eqref{weylmetric} of Petrov type I without further specification to particular examples.

The orthonormal frame adapted to the static observers with 4-velocity $U=e_0$ is given by
\beq
e_0=e^{-\psi}\partial_t\,,\qquad
e_1=e^{\psi-\gamma}\partial_\rho\,,\qquad
e_2=e^{\psi-\gamma}\partial_z\,,\qquad
e_3=\frac{e^{\psi}}{\rho}\partial_\phi\,.
\eeq
The associated NP frame \eqref{NPvsorthon} is a transverse frame with
\bea
\frac{e^{2(\gamma-\psi)}}{2}(\psi_0-\psi_4)&=&i\left[
\psi_{,\rho z}+\rho\psi_{,z}(\psi_{,z}^2-3\psi_{,\rho}^2)+3\psi_{,z}\psi_{,\rho}
\right]
\,,\nonumber\\
\frac{e^{2(\gamma-\psi)}}{2}(\psi_0+\psi_4)&=&
\psi_{,\rho\rho}+\frac1{2\rho}\psi_{,\rho}+\frac32(\psi_{,\rho}^2-\psi_{,z}^2)-\rho\psi_{,\rho}(\psi_{,\rho}^2-3\psi_{,z}^2)
\,,\nonumber\\
-\frac{e^{2(\gamma-\psi)}}{2}\psi_2&=&
\psi_{,\rho}^2-\frac1{\rho}\psi_{,\rho}+\psi_{,z}^2
\,.
\eea
The corresponding expression \eqref{tildeMdef} for the Arianrhod-McIntosh invariant is rather involved, so we avoid showing it.
The Riemann tensor is purely electric and given by
\bea
E(U)&=&\left[-\frac12(\psi_0+\psi_4)+\psi_2\right]e_1\otimes e_1
+\left[\frac12(\psi_0+\psi_4)+\psi_2\right]e_2\otimes e_2\nonumber\\
&&
- 2\psi_2\, e_3\otimes e_3
+\frac{i}2(\psi_0-\psi_4)(e_1\otimes e_2+e_2\otimes e_1)
\,.
\eea 

A canonical frame is obtained by performing a rotation of class III, which leaves $\psi_2$ unchanged ($\psi_2'=\psi_2$), whereas $\psi_0\to \psi_0'= {\mathcal A}^{-2}\psi_0$ and $\psi_4 \to \psi_4'= {\mathcal A}^2 \psi_4$, with ${\mathcal A}^2=\sqrt{\psi_0/\psi_4}$.
In fact, in the new canonical frame $\psi_0'=\psi_4'$,  that is ${\mathcal A}^{-2}\psi_0={\mathcal A}^2 \psi_4$, which implies 
\beq
\sqrt{\frac{\psi_0}{\psi_4}}={\mathcal A}^2\,,
\eeq
as well as
\beq
\omega\equiv \frac{\psi_2'}{\psi_0'}=\frac{\psi_2}{\sqrt{\psi_4 \psi_0}}\,,
\eeq
with $\omega$ real, since both $\psi_2$ and $\psi_0\psi_4$ are real quantities.
The root \eqref{lambda1sol} then has the value
\beq
\label{lambda1weyl}
\lambda_1=\left[-3\omega-\sqrt{9\omega^2-1}\right]^{1/2}
=\sqrt{\frac{1-3\omega}{2}}-i \sqrt{\frac{1+3\omega}{2}}\,.
\eeq
Only the following distinct cases are possible: 
\begin{enumerate}
\item  
$\omega\geq\frac13$, with $\lambda_1$ purely imaginary;
\item 
$\omega\leq-\frac13$, with $\lambda_1$ real;
\item 
$-\frac13<\omega<\frac13$, with $\lambda_1$ complex, and $|\lambda_1|^2=1$.
\end{enumerate}
Each of these conditions make ${\mathcal V}=0$, implying the linear dependence of the PNDs.
This completes the proof that any type I static axisymmetric vacuum spacetime is necessarily nonmaximally spanning.
In this case one can evaluate the vector
\beq
\label{Omega123cyl}
\Omega_{123}^*=[k_1\wedge k_2 \wedge k_3]^* =-24 \sqrt{1-9\omega^2}\, e_1\,,
\eeq
that is, the normal direction to the 3-plane containing the three independent PNDs $k_1$, $k_2$, $k_3$ is spacelike and aligned with the radial direction along $\rho$, unless $\omega \not = \pm 1/3$ corresponding to $\lambda_1=-i$ (when $\omega=\frac13$) or $\lambda_1=1$ (when $\omega=-\frac13$). In both these cases $\Omega_{123}^*=0$, i.e., also $\Omega_{123}$ degenerates: $\Omega_{123}=0$, and the dimension of the span of the three PNDs $k_1$, $k_2$, $k_3$ reduces to 2 and the spacetime itself reduces to the Petrov type D.
In the non-degenerate cases, however, Eq.~\eqref{Omega123cyl} gives a special geometrical meaning to the radial direction (unnoticed before) as being directly related to the null cone structure of the spacetime \eqref{weylmetric}.

\subsection{Dunn and Tupper spacetime}

Consider the Dunn and Tupper Bianchi type VI spatially homogeneous spacetime (see Ref.~\cite{DT1} and Chapter 12 of Ref.~\cite{Stephani:2003tm} as well as Ref.~\cite{Bini:2021kjm} for a recent review)  
\begin{equation}
ds^2=dt^2-\frac{t^2}{(m-n)^2}dx^2-t^{-2(m+n)}(e^{-2x}dy^2+e^{2x}dz^2)\,,
\label{(3.4)}
\end{equation}
where  $m\not=n$ are two constant parameters.
It is an exact solution of the Einstein equations sourced by a perfect fluid
with 4-velocity $U=\partial_t$
(i.e., at rest with respect to the space coordinates), and energy density and pressure given by
\begin{equation}
\rho= \frac{m^2+mn+n^2}{t^2} ,\qquad
p = -\frac{4mn}{t^2},
\label{(3.6)}
\end{equation}
respectively, provided that $m$ and $n$ satisfy the additional constraint
\begin{equation}
m(2m+1)+n(2n+1)=0.
\label{(3.7)}
\end{equation}
Note that the conditions $\rho>0$ and $p\ge 0$ require $mn\le 0$. 
The strong energy conditions
$\rho+p\ge 0$ and  $\rho+3p\ge 0$ are always satisfied for this family.
The special case of a dust fluid (i.e., with $p=0$) corresponds to either $m = 0$ or $n = 0$, but not both simultaneously zero since the resulting vacuum spacetime is flat. 

The timelike unit vector $U=\partial_t\equiv e_0$ is completed to an adapted orthonormal frame by normalizing the spatial coordinate frame
\beq
\label{frame_u}
e_1=\frac{(m-n)}{t}\partial_x
\,,\quad
e_2=e^x t^{m+n}\partial_y\,,\quad
e_3=e^{-x} t^{m+n}\partial_z.
\eeq
The electric and magnetic parts of the Weyl tensor expressed in this frame are given respectively by
\begin{eqnarray}
E(U) 
&=&\frac{(m-n)^2}{3t^2} [2e_1\otimes e_1-e_2\otimes e_2-e_3\otimes e_3]\,,
\nonumber\\
H(U) 
&=&\frac{(m-n)(m+n+1)}{t^2} [e_3\otimes e_2+e_2\otimes e_3]\,,
\end{eqnarray}

The associated NP frame \eqref{NPvsorthon} is a transverse frame with nonvanishing Weyl scalars 
\begin{eqnarray}
\psi_0&=&-\psi_4=-\frac{(m+n+1)(m-n)}{t^2},
\nonumber\\
\psi_2&=& -\frac{(m-n)^2}{t^2}.
\end{eqnarray}
Therefore, the metric \eqref{(3.4)} is of Petrov type I except for the special case $m=-n-1$ when it is instead of Petrov type D.
The Arianrhod-McIntosh invariant \eqref{tildeMdef} turns out to be
\beq
\tilde M= -\frac{729(m+n+1)^4}{(m-n)^2(13m+13n+16mn+9)^2}\,,
\eeq 
which is a real negative (possibly infinite) number for every allowed pairs $(m,n)$.

Passing to a canonical frame then gives the same expression for $\lambda_1$ as in Eq.~\eqref{lambda1weyl}, but with the purely imaginary  quantity
\beq
\omega=i\,\frac{|m-n|}{3|m+n+1|}\equiv i\,\xi  \neq0 \,,
\eeq
(nonvanishing since $m\neq n$), leading to the expression 
\beq
\lambda_1
=e^{-i\pi/4}\sqrt{3\xi+\sqrt{9\xi^2+1}}\,,
\eeq
which has both real and imaginary parts nonvanishing and cannot be a unit complex number since $\xi>0$. In fact
its absolute value is always greater than one, so that the general Dunn and Tupper solution is another example of a maximally spanning type I spacetime.

\section{Concluding remarks}

We have further characterized  Petrov type I spacetimes as maximally or nonmaximally spanning type I according to the geometrical criterion of the nonvanishing or vanishing of the wedge product of their four distinct PNDs,
corresponding to spanning a 4 or 3-dimensional subspace of the tangent space at each spacetime point.
This completes the Arianrhod-McIntosh classification of PND degeneracies based on the value of the scalar curvature invariant $\tilde M$, whose definition has no obvious relationship to this question.
These ideas have been illustrated concretely with simple examples of type I spacetimes which are maximally spanning (Petrov and Dunn-Tupper spacetimes) and some which are nonmaximally spanning (Kasner and Weyl class static cylindrical spacetimes), all of which allow a relatively straightforward computation of the distinct PNDs and their associated wedge products. 
The nonmaximally spanning Kasner case reveals an interesting correlation between the single collapsing spatial direction moving forwards in time and the orientation of the 3-subspace spanned by the PNDs,
while the Weyl spacetimes associate the normal to this 3-subspace with the cylindrical radial direction.
Other physical implications of this geometrical characterization of Petrov type I  spacetimes will be examined in future work.

\section*{Acknowledgments}
The authors thank the International Center for Relativistic Astrophysics Network (ICRANet) for partial support.
D.B. acknowledges sponsorship of the Italian Gruppo Nazionale per la Fisica Matematica (GNFM) of the Istituto Nazionale di Alta Matematica (INDAM).

\appendix

\section{Transformation properties of the NP curvature scalars}
\label{NProtations}

A Lorentz transformation of the orthonormal frame associated with a null tetrad transforms that null frame by a so called ``null rotation", which in turn transforms all of the various NP quantities.
The curvature scalars $I$ and $J$ are invariants under all the null rotations but 
the scalars $K$, $N$ and $L$ are only invariant under null rotations of class II. To understand their transformation under null rotations, we review how null rotations affect the NP curvature quantities.

Any null rotation of the basis vectors $l,n,m$ can be achieved by a succession of null rotations of the following types:
\begin{enumerate}
  \item null rotations of class I, leaving $l$ unchanged;
  \item null rotations of class II, leaving $n$ unchanged;
  \item null rotations of class III, leaving the directions of $l$ and $n$ unchanged and rotating $m$ by an angle $\theta$ in the $m-\bar m$ plane.
\end{enumerate}
The explicit transformations (see Eq.~53 of Ref.~\cite{Chandrasekhar:1985kt}) depend on the following six real parameters: $a$ (complex), $b$ (complex) and $\theta$ (real) and ${\mathcal A}$ (real), such that
\begin{enumerate}
  \item class I:
  \bea
&& l\to l\,,\quad m \to m+al \,,\quad \bar m\to \bar m +\bar a l\,,\nonumber\\
&& n\to n+\bar a m + a \bar m + a \bar a l\,.
\eea 
  \item class II: 
  \bea
&& n\to n\,,\quad m \to m+bn \,,\quad \bar m\to \bar m +\bar b l\,,\nonumber\\
&& l\to l+\bar b m + b \bar m + b \bar b n\,.
\eea 

  \item class III: 
  \bea
&& l\to {\mathcal A}^{-1}l\,,\quad n\to {\mathcal A}n \,,\quad  m \to e^{i\theta}m\,,\nonumber\\
&& \bar m\to e^{-i\theta}\bar m\,.
\eea 
\end{enumerate}
The resulting transformation laws for  the Weyl scalars are listed in many textbooks, for example \cite{Chandrasekhar:1985kt} . They are

\begin{enumerate}
  \item class I:
  \bea
&& \psi_0\to \psi_0 \,,\quad \psi_1\to \psi_1+\bar a \psi_0\,,\nonumber\\
&&  \psi_2\to \psi_2+2\bar a \psi_1 +\bar a^2 \psi_0\,,\nonumber\\
&& \psi_3 \to \psi_3+3\bar a \psi_2 +3 \bar a^2 \psi_1+\bar a^3 \psi_0\,,\nonumber\\
&& \psi_4 \to \psi_4+4\bar a \psi_3 +6 \bar a^2 \psi_2+4\bar a^3 \psi_1+\bar a ^4 \psi_0\,.
\eea 
  \item class II:

Same as the previous case with the exchange of $\ell$ and $n$, with $a\rightarrow b$ and
  \bea
&& \psi_0\leftrightarrow \bar \psi_4 \,,\quad \psi_1\leftrightarrow \bar \psi_3\,, \quad \psi_2\leftrightarrow \bar \psi_2\,, 
\eea 
i.e.,
\bea
&& \psi_4\to \psi_4 \,,\quad \psi_3\to \psi_3+b \psi_4\,,\nonumber\\
&&  \psi_2\to \psi_2+2b \psi_3 +b^2 \psi_4\,,\nonumber\\
&& \psi_1 \to \psi_1+3b \psi_2 +3 b^2 \psi_3+b^3 \psi_4\,,\nonumber\\
&& \psi_0 \to \psi_0+4b \psi_1 +6 b^2 \psi_2+4b^3 \psi_3+b ^4 \psi_4\,.
\eea 
  \item class III: 
  \bea
&& \psi_0\to {\mathcal A}^{-2}e^{2i\theta}\psi_0 \,,\quad \psi_1\to {\mathcal A}^{-1}e^{i\theta}\psi_1\,,\quad \psi_2\to \psi_2\,,\nonumber\\
&&   \psi_3 \to {\mathcal A} e^{-i\theta}\psi_3 \,,\quad \psi_4 \to {\mathcal A}^2 e^{-2i\theta}\psi_4\,.
\eea 
\end{enumerate}

The NP scalars $I$, $J$, $K$, $L$, $N$ given in Eqs. \eqref{Idef}, \eqref{Jdef} and \eqref{scalars}, respectively, are related to the discriminants of the quartic equation \eqref{lambda_eq} defining the PNDs.
Let us start with Eq.~\eqref{lambda_eq2}, i.e.,
\beq
\lambda^4 +a_1 \lambda^3 +a_2 \lambda^2+a_3\lambda +a_4=0\,,
\eeq
with rescaled coefficients $a_i$ given in Eq.~\eqref{aidef}.
The general solutions can be written as
\bea
\lambda_{1,2}&=&-\frac{a_1}{4}-\frac12\left[-\sqrt{y}\pm\sqrt{y-2\left(p+y+\frac{q}{\sqrt{y}}\right)}\right]
\,,\nonumber\\
\lambda_{3,4}&=&-\frac{a_1}{4}-\frac12\left[\sqrt{y}\pm\sqrt{y-2\left(p+y-\frac{q}{\sqrt{y}}\right)}\right]
\,,
\eea
where
\beq
p=a_2-\frac38a_1^2=\frac{6}{\psi_4^2}L\,,\qquad
q=a_3-\frac12a_1a_2+\frac18a_1^3=-\frac{4}{\psi_4^3}K\,,
\eeq
and $y$ is a solution of the auxiliary cubic equation
\beq
\label{eqy}
y^3+2py^2+(p^2-4r)y-q^2=0\,,
\eeq
with 
\beq
r=a_4-\frac14a_1a_3+\frac1{16}a_1^2a_2-\frac{3}{256}a_1^4\,,
\eeq
so that 
\beq
p^2-4r=\frac{4}{\psi_4^4}N\,.
\eeq
Writing the cubic equation \eqref{eqy} as
\beq
y^3+b_1y^2+b_2y+b_3=0\,,
\eeq
with coefficients
\beq
b_1=2p\,,\qquad
b_2=p^2-4r\,,\qquad
b_3=-q^2\,,
\eeq
a solution is given by
\beq
y=-\frac{b_1}{3}+\left[-\frac{Q}{2}+\sqrt{\frac{Q^2}{4}+\frac{P^3}{27}}\right]^{1/3}
+\left[-\frac{Q}{2}-\sqrt{\frac{Q^2}{4}+\frac{P^3}{27}}\right]^{1/3}
\,,
\eeq
where
\beq
P=b_2-\frac13b_1^2=-\frac{4}{\psi_4^2}I\,,\qquad
Q=b_3-\frac13b_1b_2+\frac2{27}b_1^3=\frac{16}{\psi_4^3}J\,.
\eeq

The scalars $K$, $N$ and $L$ are invariant under null rotations of class II, but transform under null rotations of class I and III, respectively, as follows:
\bea
K&\to&K+(2\psi_1\psi_4\psi_3-9\psi_4\psi_2^2+6\psi_3^2\psi_2+\psi_0\psi_4^2)\bar a\nonumber\\
&&
+5(-3\psi_1\psi_4\psi_2+2\psi_3^2\psi_1+\psi_0\psi_4\psi_3)\bar a^2\nonumber\\
&&
+10(\psi_0\psi_3^2-\psi_1^2\psi_4)\bar a^3
-5(2\psi_1^2\psi_3-3\psi_0\psi_3\psi_2+\psi_1\psi_4\psi_0)\bar a^4\nonumber\\
&&
-(\psi_0^2\psi_4+6\psi_1^2\psi_2+2\psi_1\psi_3\psi_0-9\psi_2^2\psi_0)\bar a^5
-(-3\psi_1\psi_2\psi_0+\psi_0^2\psi_3+2\psi_1^3)\bar a^6
\,,\nonumber\\
L&\to&L+(-2\psi_2\psi_3+2\psi_1\psi_4)\bar a
+(\psi_0\psi_4-3\psi_2^2+2\psi_1\psi_3)\bar a^2\nonumber\\
&&
+2(-\psi_1\psi_2+\psi_0\psi_3)\bar a^3
+(-\psi_1^2+\psi_2\psi_0)\bar a^4
\,,\nonumber\\
N&\to&12 L'^2-\psi_4'^2\,I
\,,
\eea
and
\beq
K\to {\mathcal A}^3e^{-3i\theta}K \,,\quad
L\to {\mathcal A}^2e^{-2i\theta}L \,,\quad
N\to {\mathcal A}^4e^{-4i\theta}N \,.
\eeq


\end{document}